\documentstyle[aps,amssymb]{revtex}

\begin{document}
\title{{\bf Anisotropy effects in a mixed quantum-classical Heisenberg model in two
dimensions}}
\author{C.\ Meyers, Y. Meurdesoif and Y. Leroyer,}
\address{Centre de Physique Th\'eorique et de Mod\'elisation de Bordeaux\\
Universit\'e Bordeaux I, CNRS, Unit\'e Associ\'ee 1537\\
19 rue du Solarium, 33174 Gradignan Cedex, France\\
{\rm and}}
\author{O. Kahn}
\address{Laboratoire des Sciences Mol\'{e}culaires\\
Institut de Chimie de la Mati\`ere Condens\'ee de Bordeaux\\
CNRS, UPR 9048\\
Avenue Albert Schweitzer, 33608 Pessac Cedex, France}
\date{24/06/97}
\maketitle

\begin{abstract}
We analyse a specific two dimensional mixed spin Heisenberg model with
exchange anisotropy, by means of high temperature expansions and Monte Carlo
simulations. The goal is to describe the magnetic properties of the compound
(NBu$_{4})_{2}$Mn$_{2}$[Cu(opba)]$_{3}\cdot $6DMSO$\cdot $H$_{2}$O which
exhibits a ferromagnetic transition at $T_{c}=15K$. Extrapolating our
analysis on the basis of renormalisation group arguments, we find that this
transition may result from a very weak anisotropy effect.
\end{abstract}

\section{
\protect\footnotetext{
CPTMB-97-9}Introduction.}

In the last few years there has been increasing interest in magnetic systems
of low dimensionality. For example, the rapidly developing field of
molecular magnetism\cite{Kahn-book} deals mainly with quasi one-dimensional
and quasi two-dimensional compounds. Although the basic theory of their
magnetic properties has been known for a long time\cite{Heis} it is now
necessary to apply it to the various contexts corresponding to these complex
molecular architectures.

The compound (NBu$_{4})_{2}$Mn$_{2}$[Cu(opba)]$_{3}\cdot $6DMSO$\cdot $H$%
_{2} $O, synthesized by H.O. Stumpf {\em et al}\cite{Kahn}, exhibits a
transition at $T_{c}=15$K towards a ferromagnetically ordered state. The
structure of this material can be schematically described by a superposition
of negatively charged layers of hexagonal lattices with the Mn$^{II}$ ions
(spin 5/2) occupying the vertices and the Cu$^{II}$ ions (spin 1/2)
occupying the middle of the links (see figure 1). The tetrabutylamonium
cations, NBu$_{4}^{+}$, are located between the layers.\ Other compounds of
the same kind have been synthesized, differing by the nature of the cations
between the layers\cite{Kahn-3}. When these cations are small,(Na$^{+}$, K$%
^{+}$ and tetramethylammonium) a long range antiferromagnetic ordering in
observed in zero field. An external field of the order of 0.15 kOe is
sufficient to overcome the very weak interlayer interactions and to lead to
a ferromagnetic-like state. The compounds then behave as metamagnets. When
the cations are larger (tetraethylammonium and beyond), a ferromagnetic
ordering occurs at a critical temperature $T_{c}$. The value of this
critical temperature first remains constant and equal to 15K, then decreases
very smoothly as the cation size increases.\ In other respects, replacing Mn$%
^{II}$ by a more anisotropic spin carrier such as Co$^{II}$ results in a
significant increase of $T_{c}$. These results suggest that both interlayer
interactions and spin anisotropy are involved in the mechanism of long range
ordering. The role of the spin anisotropy in the magnetic properties of
two-dimensional compounds is much less documented than the three dimensional
effects, and the goal of this paper is to address this problem.

In the layer of (NBu$_{4})_{2}$Mn$_{2}$[Cu(opba)]$_{3}\cdot $6DMSO$\cdot $H$%
_{2}$O, the nearest neighbour Mn$^{II}$ and Cu$^{II}$ ions interact through
an antiferromagnetic coupling. The interlayer interaction in any case is
very small as compared to the intralayer one, so that, to a good
approximation, the spin system can be considered two-dimensional. In a
previous paper\cite{Leandri}, we have shown that such a simple description,
in which the 5/2\ spins Mn$^{II}$ spins are approximated by classical ones,
gives a good account of the magnetic and thermal properties of the
paramagnetic phase of the compound. However, since the isotropic $O(3)$
model is critical only at zero temperature\cite{MerminW} we must include a
symmetry breaking mechanism in order to explain the phase transition at $%
T_{c}=15K.$ We attribute this symmetry breaking to the presence of spin
anisotropy.

Since no single crystal of the Cu-Mn compound has been obtained so far, a
direct measurement of the orientation of the anisotropy is not possible.\
However, the existence of a spontaneous magnetization below $T_{c}$ is the
signal of an axial anisotropy and of an Ising-like transition.\ An in-plane
anisotropy would have driven the system to be described by an XY symmetric
model, which, in two dimensions, exhibits a Kosterlitz-Thouless transition
with no ordered phase\cite{KT}.

The aim of this paper, is to investigate the effect of a small axial
anisotropy in the simple model described above.

For weak anisotropy, the critical properties of the model are dominated by
the cross-over between the $T=0$ critical point of the 2D Heisenberg model
and the Ising one at $T_{c}$. This effect has been widely analysed in the
framework of the purely classical Heisenberg model\cite
{Jasnow,Pfeuty,Binder,Nels-Pelc,Anis,Cucco}. In particular, the
renormalisation group analysis\cite{Nels-Pelc} leads to the following
result~: if $\lambda $ is the anisotropy parameter ($\lambda =0$ corresponds
to the isotropic case) the Ising critical temperature decreases to zero as 
\begin{equation}
T_{c}(\lambda )\approx \frac{1}{\left| \ln \lambda \right| }\quad \text{for }%
\lambda \rightarrow 0  \label{Tc}
\end{equation}
Furthermore, the zero-field susceptibility satisfies a scaling law 
\begin{equation}
\chi (\lambda ,T)=\chi (0,T)\;\Phi (\lambda e^{\frac{4\pi }{T}})  \label{RG0}
\end{equation}
where the function $\Phi (x)\approx \left| x-x_{c}\right| ^{-7/4}$ for $%
x\approx x_{c}=\lambda e^{\frac{4\pi }{T_{c}}}$. This equation gives
additional information on the Ising critical region. Let us define the width
of this region, $\delta ,$ by 
\[
1-\delta \leq \frac{T}{T_{c}}\leq 1+\delta \Longrightarrow \chi (\lambda
,T)\gg \chi (\lambda =0,T) 
\]
From eq.(\ref{RG0}) we get $\delta \approx \frac{1}{\left| \ln \lambda
\right| }$ when $\lambda \rightarrow 0$. Therefore in the weak anisotropy
limit we expect the Ising critical region to become very narrow and quite
close to $T=0$. Clearly, this makes experimental investigation difficult.

On the basis of universality, we transpose these renormalisation group
results to our mixed spins system. Since we expect the anisotropy to be very
small, we need to develop methods specifically designed to handle this
cross-over effect. In section II we present our techniques of high
temperature expansions on the one hand and of Monte-Carlo simulations on the
other hand. In section III we analyse our results and, from a comparison
with the experimental data, we determine the value of the anisotropy for the
Cu-Mn compound. Conclusions are drawn in the last section.

\section{The model.}

We denote by ${\bf S}_{j}^{{\rm (Mn)}}$ the spin $\frac{5}{2}$ operator
associated with the Mn ion at site $j$, and by ${\bf S}_{i}^{{\rm (Cu)}}$
the spin $\frac{1}{2}$ operator corresponding to the Cu ion at site $i$ in
the middle of a link of the honeycomb lattice. The antiferromagnetic
interaction is represented by the Heisenberg hamiltonian 
\begin{equation}
{\cal H}=J\sum_{<i,j>}\left( {\bf S}_{i}^{{\rm (Cu)}}\cdot {\bf S}_{j}^{{\rm %
(Mn)}}+\lambda S_{i}^{z{\rm (Cu)}}S_{j}^{z{\rm (Mn)}}\right) -\left(
\sum_{j=1}^{N_{S}}g_{1}\mu _{B}{\bf S}_{j}^{{\rm (Mn)}}+%
\sum_{i=1}^{N_{L}}g_{2}\mu _{B}{\bf S}_{i}^{{\rm (Cu)}}\right) \cdot {\bf H}
\label{Heisenberg}
\end{equation}
where $J$ is positive, $\lambda (>0)$ is the anisotropy parameter, ${\bf H}$
is the external magnetic field, $<i,j>$ stands for a pair of nearest
neighbour spins, $N_{S}$ is the number of sites and $N_{L}$ is the number of
links on the honeycomb lattice ($N_{L}=3/2N_{S}$). The spin $\frac{5}{2}$
operator can be approximated by a {\em classical} spin $S.{\bf s}$ where $%
{\bf s}$ is a unit classical vector and $S=\sqrt{\frac{5}{2}(\frac{5}{2}+1)}$%
, whereas the spin $\frac{1}{2}$ operators are expressed in terms of the
Pauli matrices, ${\bf S}^{{\rm (Cu)}}=\frac{1}{2}{\bf \sigma }$. Since the
quantum spin sites are not directly coupled to each other, one can trace out
the quantum spin dependence to get a completely classical partition function 
\begin{equation}
Z(T,{\bf H})=\int \left( \prod_{i=1}^{N_{S}}d\Omega _{i}\right) ~\left\{
\prod_{<ij>}2\cosh \left\| {\bf W}_{ij}+\frac{1}{2}\beta g_{2}\mu _{B}{\bf H}%
\right\| \right\} ~\exp \left( \beta g_{1}\mu _{B}S\,{\bf H\cdot }%
\sum_{i=1}^{N_{S}}{\bf s}_{i}\right)  \label{Z}
\end{equation}
where we have defined 
\begin{equation}
{\bf W}_{ij}=-\frac{1}{2}\beta JS\left( {\bf s}_{i}+{\bf s}_{j}+\lambda
(s_{i}^{z}+s_{j}^{z})\widehat{{\bf e}}_{z}\right)  \label{Wij}
\end{equation}
and $\left\| {\bf X}\right\| $ stands for the length of the vector ${\bf X}$%
. The indices $i$ and $j$ now label the {\em classical} spins located at the
vertices of the honeycomb lattice.

By choosing the orientation of the magnetic field parallel to the anisotropy
axis, or orthogonal to it (along the $x$ axis), we define $Z_{\mu
}(T,H)\equiv Z(T,H\,\widehat{{\bf e}}_{\mu })$ with $\mu =x$ or $z$, as
follows~: 
\begin{equation}
Z_{\mu }(T,H)=\int \left( \prod_{i=1}^{N_{S}}d\Omega _{i}\right) ~\left\{
\prod_{<ij>}2\cosh \left( K\,\phi _{<ij>}^{\mu }\right) \right\} ~\exp
\left( \beta g_{1}\mu _{B}SH\;\sum_{i=1}^{N_{S}}s_{i}^{\mu }\right) \qquad
\mu =x\text{ or }z  \label{Zmu}
\end{equation}
with $K=\frac{1}{2}\beta JS$ and $\phi _{<ij>}^{\mu }=\left\| {\bf s}_{i}+%
{\bf s}_{j}+\lambda (s_{i}^{z}+s_{j}^{z})\widehat{{\bf e}}_{z}-\frac{%
g_{2}\mu _{B}}{JS}H\,\widehat{{\bf e}}_{\mu }\right\| $

We shall be interested in the standard observables : the specific heat $%
C_{V}=k_{B}T^{2}\frac{\partial ^{2}}{\partial T^{2}}\ln Z(T,0),$ the
suceptibility along the different directions ${\chi }_{\mu }{=\frac{k_{B}T}{V%
}}\left. {\frac{\partial ^{2}}{\partial H^{2}}\ln Z}_{\mu }\right| _{H=0}$,
and the total susceptibility, ${\chi }=\frac{1}{3}\,{\chi }_{z}+\frac{2}{3}\,%
{\chi }_{x}$, which is measured experimentally

\section{The method of analysis}

\subsection{The high temperature expansion.}

We have performed the expansion of $\ln Z_{\mu }$ in power series of $K$ up
to the 19th order and to the second order in $H$ for computing the magnetic
susceptibility. Then we analysed the series for decreasing values of the
anisotropy parameter. The complexity of the nearest neighbour interaction
(eq.(\ref{Zmu})) does not allow the standard techniques\cite{Jasnow,DB,McK}
to be used.

The diagrammatic expansion is generated by replacing in eq.(\ref{Zmu}) each $%
\cosh \left( K\,\phi _{<ij>}^{\mu }\right) $ term by $1+\Psi _{<ij>}^{\mu }$
where the function $\Psi _{<ij>}^{\mu }$ results from the expansion of the
hyperbolic cosine in power series of $K$ to the given maximal order and of $%
H $ to the second order. To each function $\Psi _{<ij>}^{\mu }$ appearing in
the product over the nearest neighbour pairs in eq.(\ref{Zmu}), is
associated one link of a graph $G$. In this way, a link contains all
positive powers of $K$.

The powerfull star graph expansion technique\cite{McK} cannot be used here
since,\ due to the presence of the anisotropy term, the partition function
of articulated graphs does not factorise. Our procedure is based on the the
standard connected graph expansion\cite{McK} for the normalised partition
function $\widetilde{Z}_{\mu }=Z_{\mu }(T,H)/\left. Z_{\mu }(T,H)\right|
_{J=0}$ 
\[
\ln \widetilde{Z}_{\mu }(T,H)=\sum_{\{G\}}C(G)\,\omega (G) 
\]
where $\{G\}$ is the set of all connected graphs to a given order, $C(G)$
the embedding cons$\tan $ts of the graph $G$, and $\omega (G)$ its weight.
The weights $\omega (G)$ are constructed through the recursive technique $%
\ln \widetilde{Z}_{\mu }(G)=\sum_{\{g\}}\omega (g)$, where $\{g\}$ is the
set of all subgraphs of $G$.

The main difficulty of the method resides in the computation of $\ln 
\widetilde{Z}_{\mu }(G),$ given by : 
\begin{equation}
\ln \widetilde{Z}_{\mu }(G)=\int \left( \prod_{i\in G}d\Omega _{i}\right)
\left( \prod_{\ell \in G}\Psi _{\ell }^{\mu }\right) ~\exp \left( \beta
g_{1}\mu _{B}SH\sum_{i\in G}s_{i}^{\mu }\right)  \label{lnZ(G)}
\end{equation}
where $\{\ell \}$ and $\{i\}$ are respectively the set of links and of
vertices belonging to the graph $G$.\ We proceed as follows :

\begin{enumerate}
\begin{enumerate}
\item  By using the spherical harmonic basis and the reduction formula, each
function $\Psi _{<ij>}^{\mu }$ can be expressed as~: 
\[
\Psi _{<ij>}^{\mu }=\sum_{l_{1}m_{1}l_{2}m_{2}}\Lambda
_{l_{1}m_{1}}^{l_{2}m_{2}}\;Y_{l_{1}m_{1}}(\Omega
_{i})\;Y_{l_{2}m_{2}}(\Omega _{j}) 
\]
where $\Lambda _{l_{1}m_{1}}^{l_{2}m_{2}}$ is a matrix built recursively of
which elements are power series in $K$ and $H$.

\item  The exponential term in eq.(\ref{lnZ(G)}) is expanded to second order
in $H,$ and the spin dependence expressed in terms of the spherical
harmonics.

\item  The integral is then computed, contracting the products of spherical
harmonics by means of the reduction formula, and then integrating over the
residual angular variables.
\end{enumerate}
\end{enumerate}

As an illustration of our results, we give in table I the coefficients of
the development of $\chi _{z}$ to the zero$^{th}$ and first order in $%
\lambda $.

The series for the zero-field susceptibility is then analysed by means of
the Pad\'{e} extrapolation technique. For our system, the renormalisation
group result of eq.(\ref{RG0}) becomes\cite{Leandri} : 
\begin{equation}
\frac{\chi (\lambda ,K)}{\chi (0,K)}=\Phi (\lambda e^{\frac{2\pi }{\sqrt{3}}%
K})\qquad \text{with }K=\frac{1}{2}\beta JS  \label{RG}
\end{equation}
where the function $\Phi (x)\approx \left| x-x_{c}\right| ^{-7/4}$ for $%
x\approx x_{c}=\lambda e^{\frac{2\pi }{\sqrt{3}}K_{c}}$. For small $\lambda $%
, the singularity is located at large values of $K$, far away from the
perturbative region. According to these results, we analysed the function $%
\left( \frac{\chi (\lambda ,K)}{\chi (0,K)}\right) ^{4/7}$ in terms of the
variable $v=1-e^{-\alpha K},$ small at high temperature but bounded for
large $K$.\ The variational parameter $\alpha $ is optimised by stabilising
the Pad\'{e} table. For very low values of $\lambda $ the determination of
the critical temperature resulting from the location of the poles becomes
inaccurate giving large error bars.\ For $\lambda \lesssim 10^{-2}$ an
extension of the series to higher orders is needed to reliably get precise
results from this method. However, in this very weak regime of anisotropy,
once the critical temperature is known by another method - Monte-Carlo
simulation for instance - the series becomes a useful analytic
representation of the observables. It will be used to determine the physical
parameters from a comparison with the experimental data.

The results of this analysis will be presented and discussed in the next
section together with the Monte-Carlo results.

\subsection{The Monte-Carlo simulation.}

We performed a Monte Carlo simulation in order to verify the results of the
high temperature expansion, and to investigate the very small anisotropy
regime where the perturbative technique fails. The simulation is based on
the effective classical model 
\begin{equation}
-\beta {\cal H}_{\text{eff}}=\sum_{<ij>}\ln \left( 2\cosh \left\| {\bf W}%
_{ij}\right\| \right)  \label{Heff}
\end{equation}
where ${\bf W}_{ij}$ is defined in eq.(\ref{Wij}). The various observables
can be expressed as ensemble averages with respect to the Boltzmann weight ${%
e^{-\beta ~{\cal H}_{{\rm eff}}}/{Z(T,0)}}$. Their expression is a simple
generalisation of the definition of eqs.(6,8) of ref.\cite{Leandri} in which 
$W_{ij}$ is replaced by the the new definition of eq.(\ref{Wij}).

Our goal is to explore the weak anisotropy regime where the cross over
effect between the 2D Heisenberg and Ising fixed points is important. As the
anisotopy gets smaller, most of the low temperature region is dominated by
the 2D Heisenberg regime in which the correlation length remains large due
to the essential singularity at $T=0$. Therefore, the usual limitations of
the Monte-Carlo procedure - critical slowing down and the finite size
effects - constrain rather severely the simulation of these systems.

In order to overcome the first problem we used a global algorithm.We have
adapted the Wolf algorithm\cite{Wolf} to the case of an anisotropic
interaction. The 'Ising' orientation of the spins, used to contruct the Wolf
cluster and which is randomly chosen in the standard algorithm, is imposed
here by the anisotropy. In our procedure, a Monte-Carlo update proceeds in
three steps :

\begin{enumerate}
\item  construct a first cluster with respect to the $z$ (anisotropy) axis
and flip the corresponding spin components ;

\item  construct another cluster relative to a randomly chosen direction in
the $x-y$ plane and flip the corresponding spin components ;
\end{enumerate}

\noindent After these two steps, a given spin remains on the cone defined by
its initial orientation and the $z$-axis. Therefore we proceed to another
step :

\begin{enumerate}
\item[3.]  change the actual orientation of each spin of the lattice
according to a standard Metropolis algorithm.\ This operation is repeated
twice.
\end{enumerate}

We checked that this procedure allows us to recover the results of the
zero-anisotopy case\cite{Leandri}, and of the strong anisotopy (classical
Ising) limit with a very small critical slowing down effect.

The second problem concerns the finite size effects.\ For small $\lambda $,
since we investigate the low temperature region where the correlation length
remains large even outside the Ising critical region, we need large
lattices.\ We have limited our analysis to $\lambda =0.001$, and to a
maximum lattice size of $L=256$.

\section{Results}

\subsection{The critical temperature.}

To validate our methods we compare the results obtained by both procedures.
We present in figure 2 the zero-field susceptibility, orthogonal to the
anisotropy axis ($\chi _{x}$) and parallel to it ($\chi _{z}$) for the high
temperature series and the Monte-Carlo data as a function of $K$ and for a
moderatedly small value of the anisotropy parameter $\lambda =0.1$. The
agreement between these two results is very good. These curves clearly show
the divergent behaviour of the axial susceptibility, which will be used to
determine the critical temperature in the Monte-Carlo simulation. Figure 3
displays the specific heat as a function of $K$ for the same anisotropy.
Besides the good agreement between the two methods, we observe the clear
critical signal at $K_{c}\simeq 2.21$ which emerges from the comparison with
the result for the isotropic model.

The critical temperature is obtained from the Monte-Carlo data by localising
the peak in the specific heat and the inflexion point of the susceptibility
as a function of the temperature. In order to estimate the precision of such
a determination we performed a complete finite size scaling analysis of the
data at $\lambda =0.1$ to obtain the critical temperature and the
susceptibility exponent. The results are presented in figure 4 where we
plotted $T_{c}(L),$ measured from the two signals, for several lattice sizes
(fig.(4a)), and in fig.(4b) $\ln \chi $ as a function of $\ln \left|
K_{c}-K\right| $. Fig.(4a) shows a small variation of the finite system
critical signal, which allows us to estimate the bulk critical temperature $%
T_{c}$ from lattices of size not exceeding $L=128$. With this estimate, we
determine the exponent $\gamma $ of the susceptibilty from fig.(4b).\ The
result $\gamma =1.74(2)$ is in very good agreement with the expected exact
Ising value $\gamma =1.75$. This is a self-consistent indication of the
reliability of the critical temperature measurement. Furthermore, we check
that the Monte-Carlo result falls inside the error bar obtained from the
high temperature series analysis.

With this method, we determined the critical coupling $K_{c}=\frac{1}{2}%
\frac{JS}{k_{B}T_{c}}$ as a function of the anisotropy parameter on lattices
of size $L\leq 256$ down to $\lambda =0.001$. The results of both analyses -
Monte-Carlo and high temperature expansion - are displayed in figure 5.
Fig.(5a) shows the general trend of the variations of $K_{c}^{-1}\propto
T_{c}$ as a function of $\lambda ,$ with a decrease to zero for $\lambda
\rightarrow 0$, and a linear variation at large $\lambda $. Actually, for $%
\lambda \rightarrow \infty $, the model coincides with a classical Ising
model with coupling $K_{\text{Ising}}=\lambda K$ so that, in this limit, $%
K_{c}^{-1}(\lambda )\simeq \lambda K_{c\text{ Ising}}^{-1}.$ From the high
temperature expansion, we get the estimation $K_{c\text{ Ising}}\simeq
1.46(1)$. The variations for small $\lambda $ are presented in fig.(5b)
where we plotted $K_{c}$ vs $\ln \lambda $. It appears that the behaviour
predicted by the renormalisation group (eq.(\ref{Tc})) is obtained for very
small anisotropy. Actually, from fig.(5b) we obtain :

\begin{mathletters}
\begin{equation}
K_{c}=\frac{\sqrt{3}}{2\pi }\left| \ln \lambda \right| +2.41\qquad \lambda
\lesssim 0.001.  \label{Kc}
\end{equation}
Therefore, by extrapolating this behaviour down to $\lambda \rightarrow 0$,
we are able to predict the critical temperature for very small anisotropy.

\subsection{Comparison with the experimental results.}

In order to compare with the experimental results, we proceed in two steps.

\begin{itemize}
\item  We get an estimation of the anisotropy parameter from the value of $J$
previously determined in the analysis of the paramagnetic phase\cite{Leandri}
and from the experimental critical temperature $T_{c}=15$ K. We obtain $%
K_{c}\simeq 4.6$ which corresponds to $\lambda \simeq .0004$, according to
the fig.(5b).

\item  We perform an adjustment of $J$, $g_{1}$ and $g_{2}$ at fixed $%
\lambda $ by fiting the experimental data with a selected Pad\'{e}
approximant. Small variations around the fixed $\lambda $ value do not
significantly change the fit.
\end{itemize}

The results are presented in figure 6, for the whole range of temperature in
fig.(6a) and for the critical region only in fig.(6b). An excellent
agreement with the experimental data is obtained with the following set of
parameters : 
\end{mathletters}
\[
J=45.5\text{ K}\quad ,\quad g_{1}=2.0\quad ,\quad g_{2}=2.11\quad ,\quad
\lambda \simeq .0005 
\]
The exponential variation of $\lambda $ with respect to $T_{c}$ (eq.(\ref{Kc}%
)) induces rather large error bar on the determination of $\lambda $ :
actually, an error of one kelvin on $T_{c}$ induces a variation of $\lambda $
by a factor of 3. However, the magnetic susceptibility is rather insensitive
to these variations which only affect the Ising critical region, in such a
narrow range of temperature that it is not visible experimentally. Moreover,
the high temperature ($T\geq 50$ K) behaviour is left completely unchanged
by such a small perturbation, thus preserving the good agreement with the
isotropic model observed in ref.(\cite{Leandri}).

\section{Conclusions}

We have determined the spin anisotropy which is present in the Cu-Mn
magnetic compound\cite{Kahn} and which is responsible for a ferromagnetic
transition at 15K. Two methods are used in a complementary way, high
temperature expansion and Monte-Carlo simulation, in order to extract
reliable results out of the strong cross-over regime where the effect lies.
Assuming universality and extending the renormalisation group results to our
situation, we obtain a very small value for the anisotropy parameter.

The strong behaviour $T_{c}(\lambda )\approx \frac{1}{\left| \ln \lambda
\right| }$ is responsible for the fact that a very weak perturbation ($%
\lambda \simeq 10^{-4}$) produces a sizeable effect ($T_{c}\simeq 15$ K).
For these quasi-two dimensional molecular compounds, involving high spin
magnetic ions, it is unconceivable that at such a low level of magnitude the
anisotropy is absent. Therefore we should always expect a ferromagnetic
transition at an appreciable critical temperature in these systems.

In this work, we only considered the exchange anisotropy as a source of the
O(3) symmetry breaking. Alternatively, the on site anisotropy could be
present, but in the limit of weak anisotropy under consideration here, on
the ground of universality we expect the results to be unchanged. Concerning
the interlayer interactions which have been neglected in our approach, it is
known\cite{Anis} that if the ratio of the interlayer to the intralayer
coupling, $\sigma =\frac{J_{\perp }}{J_{\parallel }}$, is small, the
transition temperature induced by these three dimensional effects behaves
like $T_{c}\approx \frac{1}{\ln \left| \sigma \right| }$.\ Therefore, even a
very small interplane coupling may significantly contribute to the measured
critical temperature, in competition with the anisotropy effect. We have
seen that increasing the interlayer distance beyond a certain limit results
in a smooth decrease of $T_{c}$. Taking into account the interlayer coupling
would require to introduce an additional cross-over effect between the three
dimensional model with spatial anisotropy and the two dimensional one with
exchange anisotropy\cite{Anis}. As a consequence, the contribution of the
spin anisotropy to the critical temperature would be overestimated and our
result for the anisotropy parameter would turn out to be an upper
bound.\bigskip

{\bf Aknowledgements}\newline
\noindent We thank S.V.\ Meshkov for his contribution to the Monte-Carlo
part of this work and J.\ Leandri for helpful discussions.

\noindent \newpage \bigskip

\begin{center}
\begin{tabular}{ccccc}
\hline\hline
$\medskip $ & \multicolumn{2}{c}{$\lambda ^{0}$} & \multicolumn{2}{c}{$%
\lambda ^{1}$} \\ \hline\hline
$\medskip $ & $S^{2}g_{1}{}^{2}$ & $g_{2}{}^{2}$ & $S^{2}g_{1}{}^{2}$ & $%
g_{2}{}^{2}$ \\ \hline
$\medskip K^{0}$ & $\frac{2}{9}$ & $\frac{{1}}{4}$ & $0$ & ${0}$ \\ 
$\medskip K^{2}$ & ${\frac{2}{9}}$ & ${\frac{2}{9}}$ & ${\frac{28\,}{45}}$ & 
${\frac{2}{3}}$ \\ 
$\medskip K^{4}$ & $0$ & $-{\frac{{1}}{15}}$ & ${\frac{32}{81}}$ & ${\frac{6%
}{25}}$ \\ 
$\medskip K^{6}$ & ${\frac{2}{225}}$ & ${\frac{533}{8505}}$ & ${\frac{1532}{%
30375}}$ & ${\frac{5282}{42525}}$ \\ 
$\medskip K^{8}$ & $-{\frac{4}{2835}}$ & $-{\frac{5683}{127575}}$ & ${\frac{%
512}{14175}}$ & $-{\frac{1754}{25515}}$ \\ 
$\medskip K^{10}$ & ${\frac{524}{893025}}$ & ${\frac{19912}{601425}}$ & $-{%
\frac{128824}{9568125}}$ & ${\frac{219628}{2525985}}$ \\ 
$\medskip K^{12}$ & $-{\frac{69464}{35083125}}$ & $-{\frac{51038503}{%
1915538625}}$ & ${\frac{117344}{526246875}}$ & $-{\frac{14083390696}{%
143665396875}}$ \\ 
$\medskip K^{14}$ & ${\frac{4539614}{1149323175}}$ & ${\frac{6380434}{%
273648375}}$ & ${\frac{5687869652}{430996190625}}$ & ${\frac{20243410804}{%
184712653125}}$ \\ 
$\medskip K^{16}$ & $-{\frac{408296024}{86199238125}}$ & ${-{\frac{%
87518482699}{4396161144375}}}$ & $-{\frac{62768262944}{4114054546875}}$ & $-{%
\frac{3168011262218}{29973825984375}}$ \\ 
&  &  &  &  \\ \hline\hline
$\medskip $ & \multicolumn{2}{c}{$\lambda ^{0}$} & \multicolumn{2}{c}{$%
\lambda ^{1}$} \\ \hline\hline
$\medskip $ & \multicolumn{2}{c}{$S\,g_{1}g_{2}$} & \multicolumn{2}{c}{$%
S\,g_{1}g_{2}$} \\ \hline
$\medskip K^{1}$ & \multicolumn{2}{c}{$-{\frac{2}{3}}$} & \multicolumn{2}{c}{%
$-{\frac{2}{3}}$} \\ 
$\medskip K^{3}$ & \multicolumn{2}{c}{$-{\frac{2\,}{27}}$} & 
\multicolumn{2}{c}{$-{\frac{166}{135}}$} \\ 
$\medskip K^{5}$ & \multicolumn{2}{c}{$-{\frac{8}{405}}$} & 
\multicolumn{2}{c}{$-{\frac{152}{675}}$} \\ 
$\medskip K^{7}$ & \multicolumn{2}{c}{${\frac{2}{8505}}$} & 
\multicolumn{2}{c}{$-{\frac{69758}{637875}}$} \\ 
$\medskip K^{9}$ & \multicolumn{2}{c}{$-{\frac{4}{4725}}$} & 
\multicolumn{2}{c}{${\frac{44524}{1913625}}$} \\ 
$\medskip K^{11}$ & \multicolumn{2}{c}{${\frac{7108}{49116375}}$} & 
\multicolumn{2}{c}{$-{\frac{316875236}{11051184375}}$} \\ 
$\medskip K^{13}$ & \multicolumn{2}{c}{${\frac{45214696}{5746615875}}$} & 
\multicolumn{2}{c}{${\frac{11621348744}{143665396875}}$} \\ 
$\medskip K^{15}$ & \multicolumn{2}{c}{$-{\frac{2434334054}{86199238125}}$}
& \multicolumn{2}{c}{$-{\frac{130568299906}{587722078125}}$} \\ 
$K^{17}$ & \multicolumn{2}{c}{${\frac{875743046944}{13987785459375}}$} & 
\multicolumn{2}{c}{${\frac{419699964912544}{887686384921875}}$}
\end{tabular}
\end{center}

{\bf Table I} : A subset of the coefficients of the high temperature series
for the axial susceptibility $\chi _{z}$ in powers of $K=\frac{1}{2}\beta JS$
up to the 17th order. The coefficient of $K^{n}$ is a polynomial of degree $%
n $ in the anisotropy parameter $\lambda $.\ For each coefficient we only
give the constant and linear term in $\lambda $\ (columns with heading $%
\lambda ^{0}$ and $\lambda ^{1}$respectively)

The top half of the table contains the even power coefficients ; for each
power of $\lambda $\ the fraction of the first sub-column must be multipied
by $S^{2}g_{1}{}^{2}$ and added to the second multiplied by $g_{2}{}^{2}$.
For instance, the coefficient of $K^{2}$ is~: $\frac{2}{9}S^{2}g_{1}{}^{2}+%
\frac{2}{9}g_{2}{}^{2}+\lambda \left( \frac{28}{45}S^{2}g_{1}{}^{2}+\frac{2}{%
3}g_{2}{}^{2}\right) +O(\lambda ^{2})$

The bottom half of the table contains the odd power coefficients ; the
fraction in each column must be multiplied by $Sg_{1}g_{2}$.

\newpage

\begin{center}
{\LARGE Figure Captions\bigskip }
\end{center}

\begin{description}
\item[{\bf Figure 1}]  : An elementary cell of the hexagonal lattice, with
the Cu ions (closed circles) at the midle of the links and the Mn ions (open
circles) at the vertices.

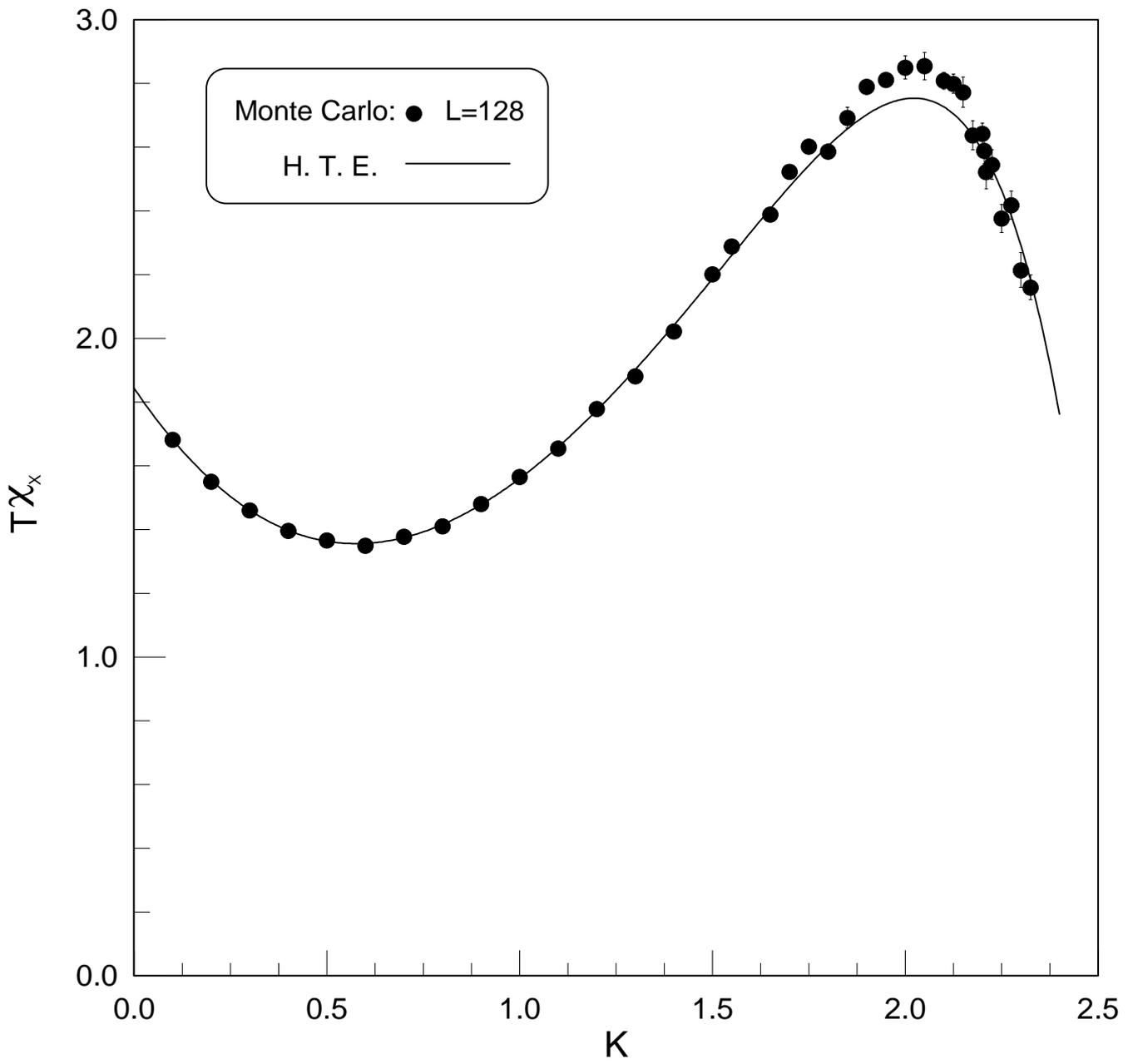
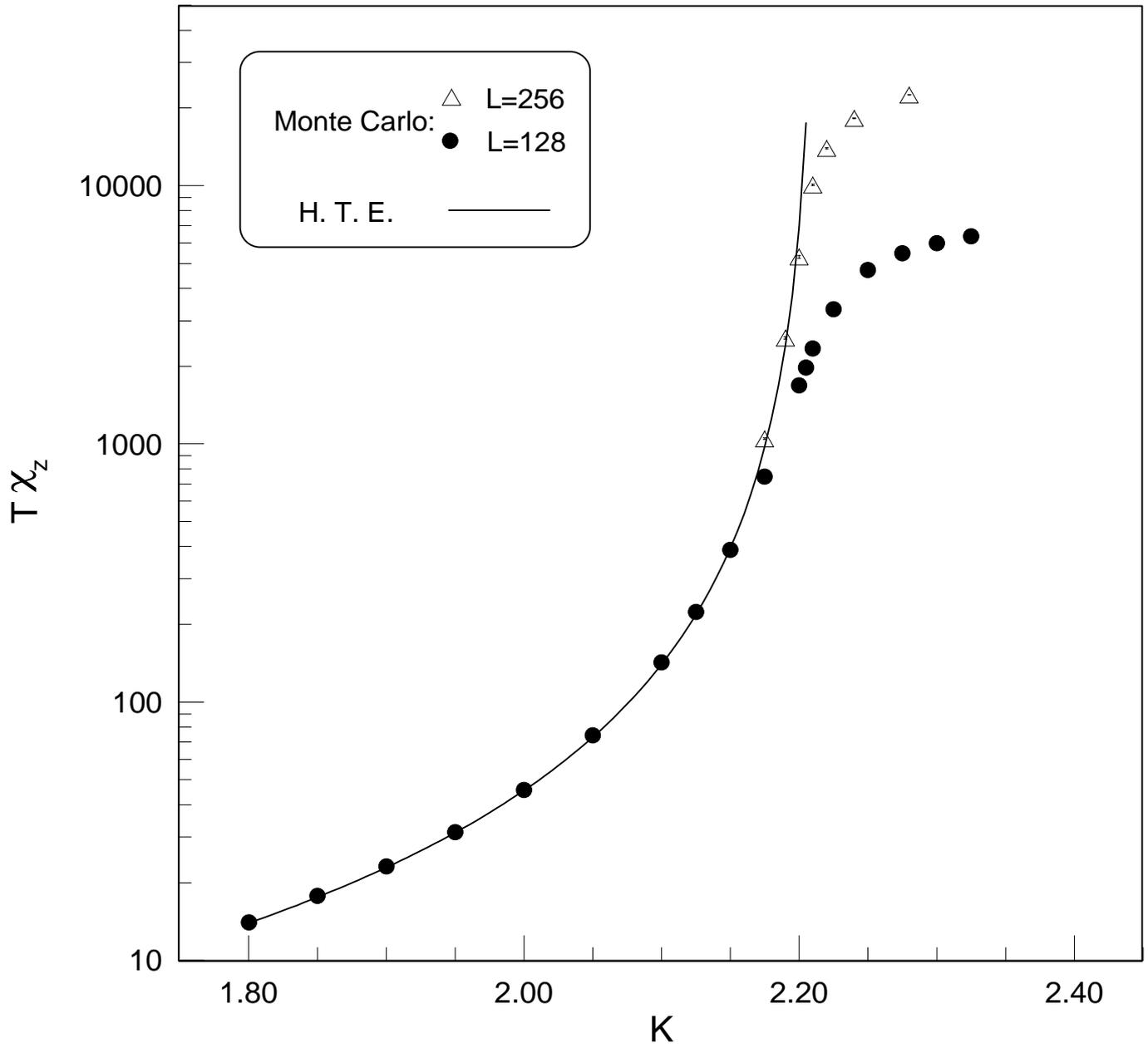
\item[{\bf Figure 2}]  : The susceptibility multiplied by the temperature,
as a function of the reduced coupling $K=\frac{1}{2}\beta JS$ for an
anisotropy $\lambda =0.1$. The data points come from the Monte-Carlo
simulation ; the solid line corresponds to the Pad\'{e} approximant of the
high temperature expansion (H.T.E.).

\begin{description}
\item[(a)]  the in-plane susceptibility ;

\item[(b)]  the axial susceptibility ; the closed circles correspond to a
lattice size $L=128$ and the triangles to $L=256$.
\end{description}

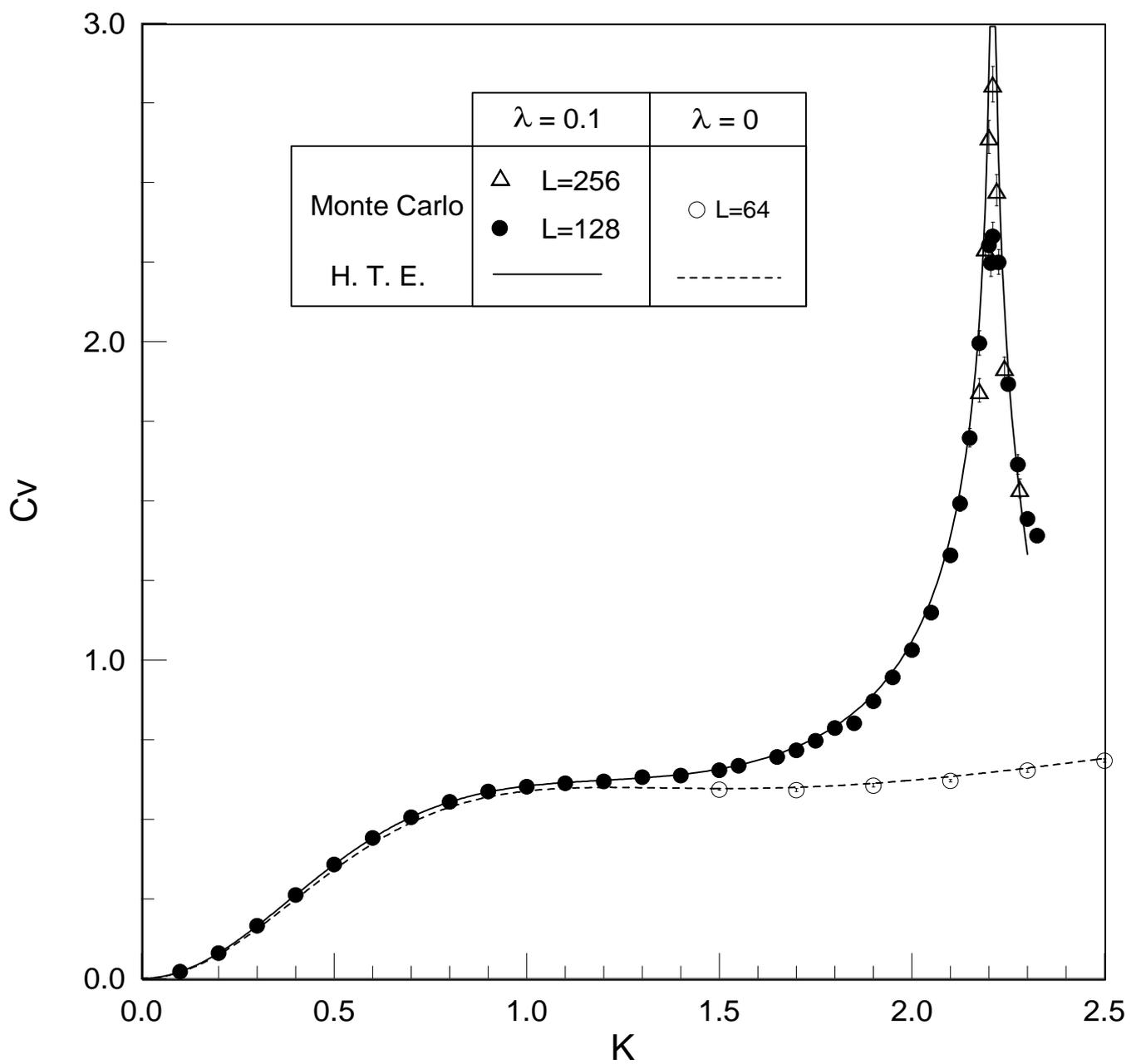
\item[{\bf Figure 3}]  : The specific heat as a function of $K$.
\end{description}

\begin{itemize}
\item  for an anisotropy parameter $\lambda =0.1$ : the Monte-Carlo data are
represented by the closed circles for a lattice size $L=128$ and by open
triangles for $L=256$. The solid line corresponds to the high temperature
expansion (Pad\'{e} approximant).

\item  for the isotropic model ($\lambda =0$) : the open circles correspond
to the Monte-Carlo simultation ($L=64$) and the dashed line to the series
result..
\end{itemize}

\begin{description}
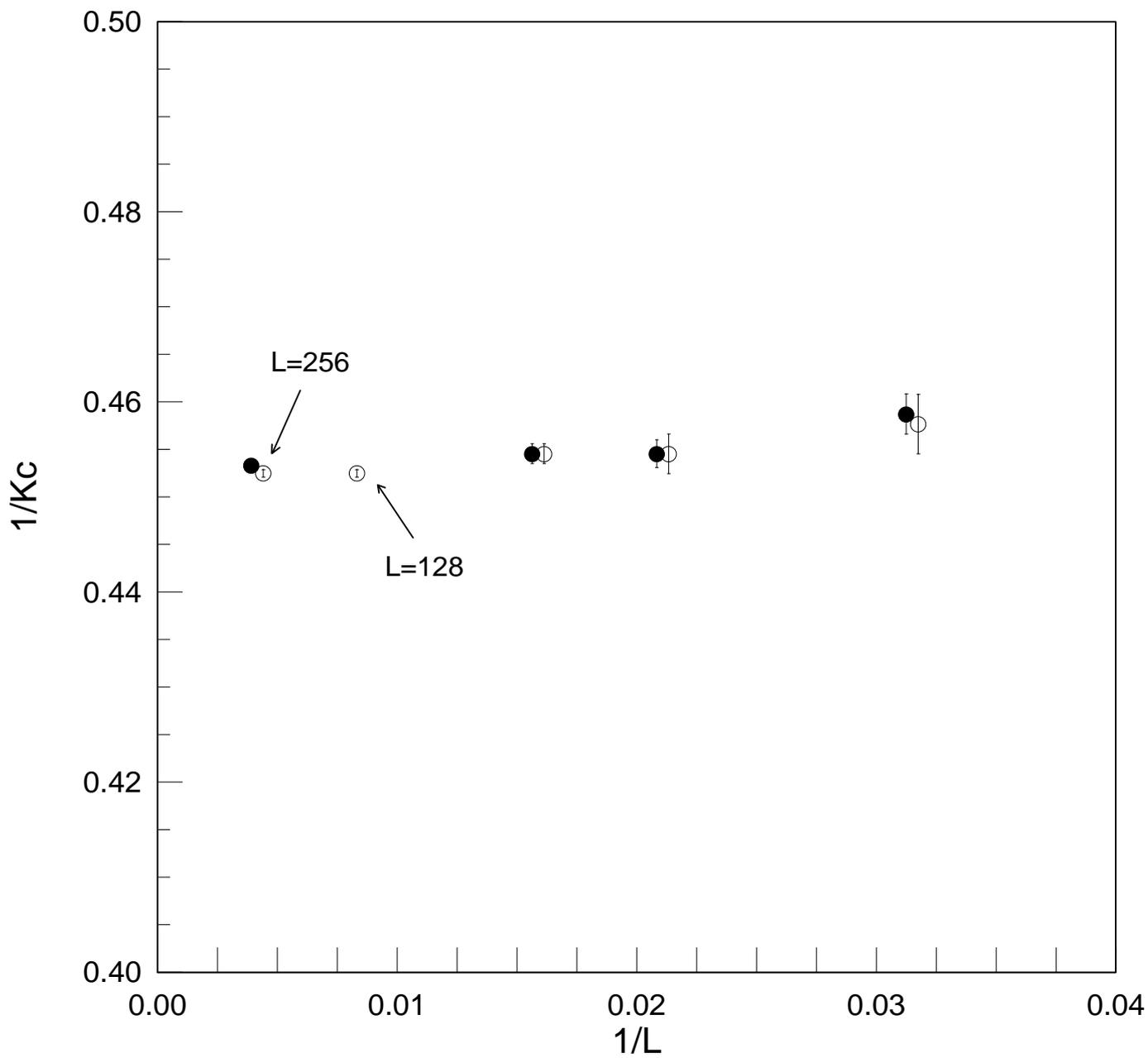
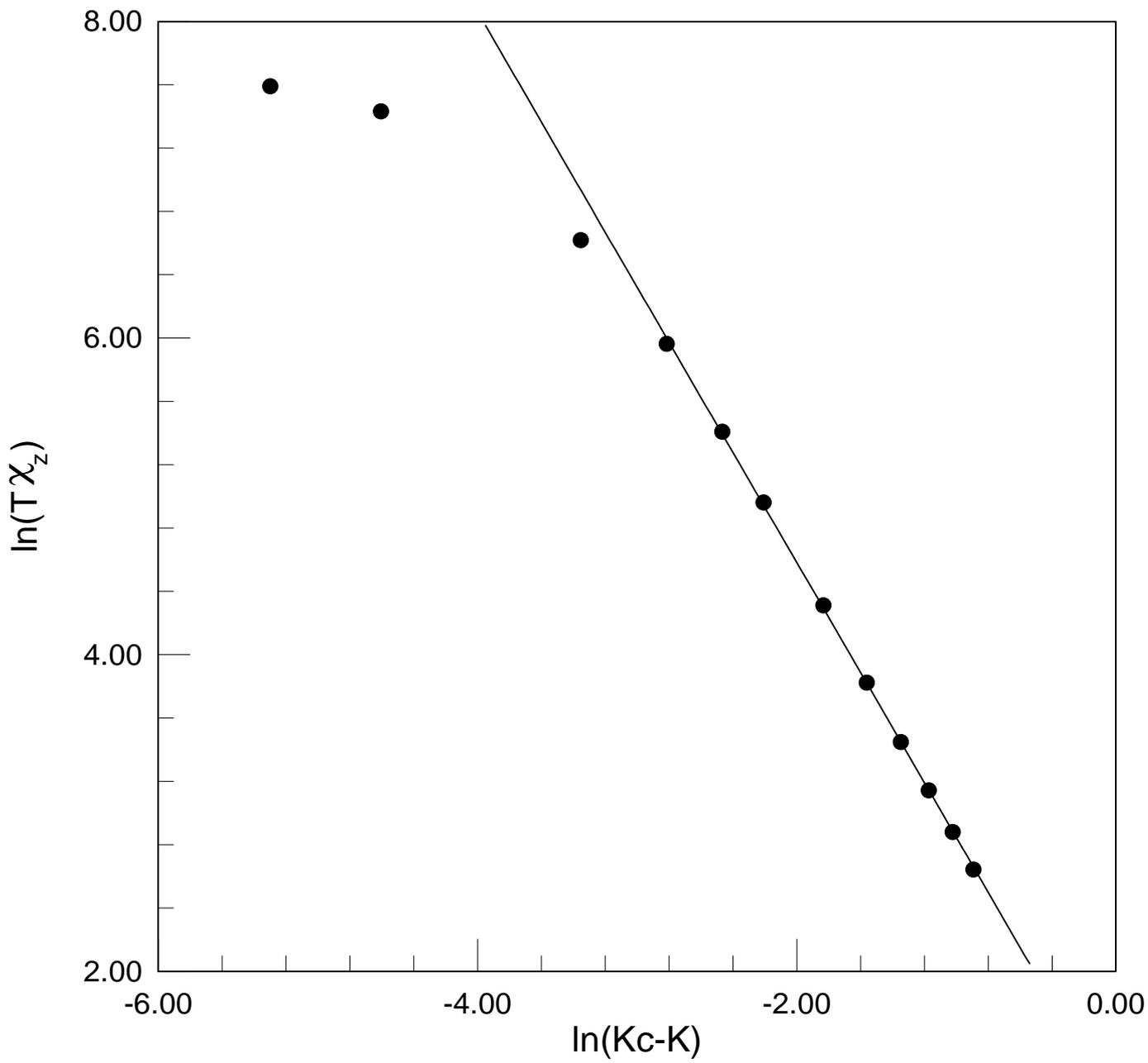
\item[{\bf Figure 4}]  Analysis of the Ising critical regime :

\begin{description}
\item[(a)]  The finite size critical temperature as a function of the
lattice size $L$ :

\begin{enumerate}
\begin{itemize}
\item  from the maximum of the specific heat (triangles)

\item  from the inflexion point of the axial susceptibility (circles)
\end{itemize}
\end{enumerate}

\item[(b)]  the axial susceptibility as as a function of $\left|
K_{c}-K\right| $ in log-log scale. The slope of the fited straight line is $%
1.74(2)$ in good agreement with the expected exact Ising value $\gamma =1.75$%
.
\end{description}

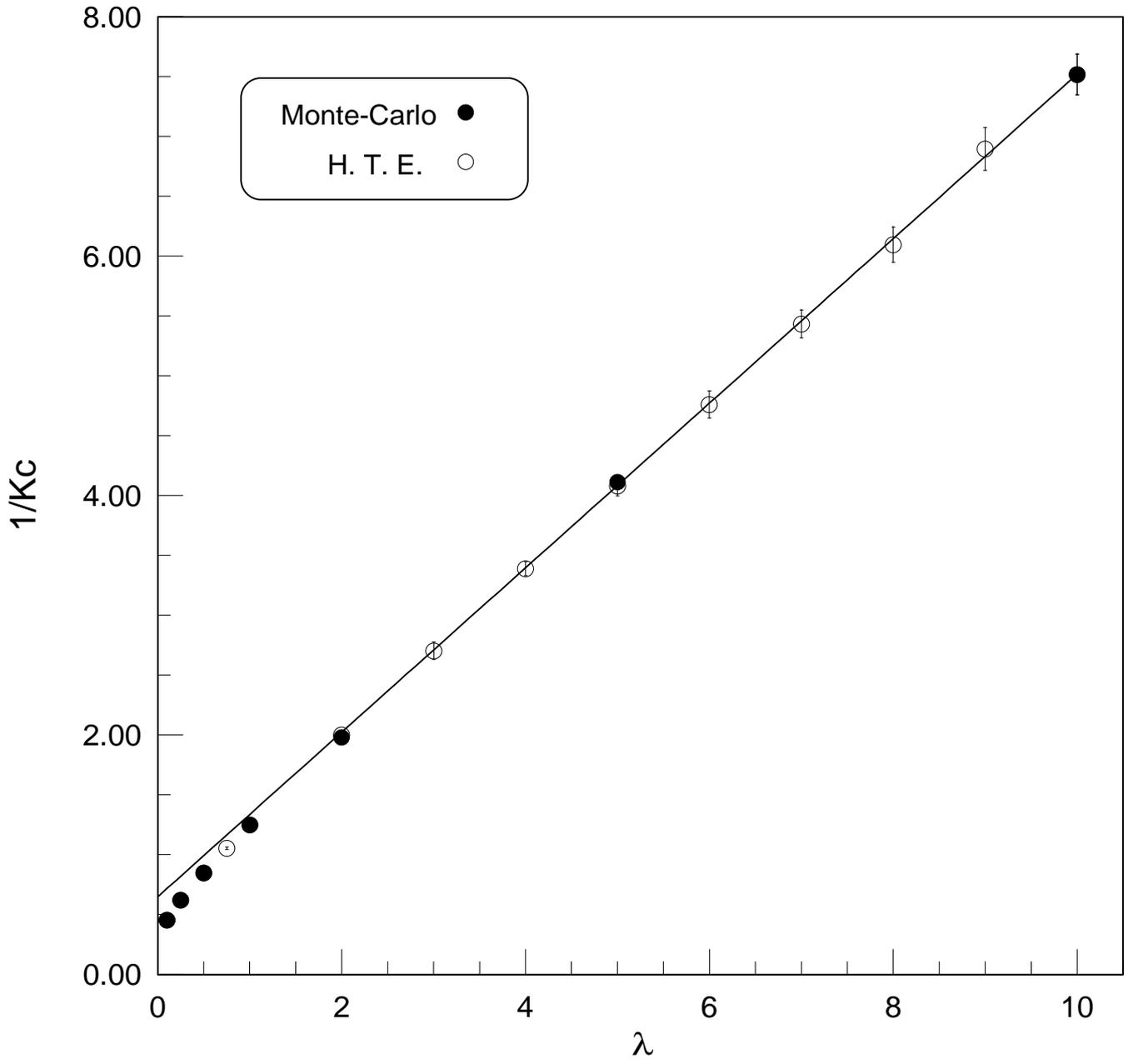
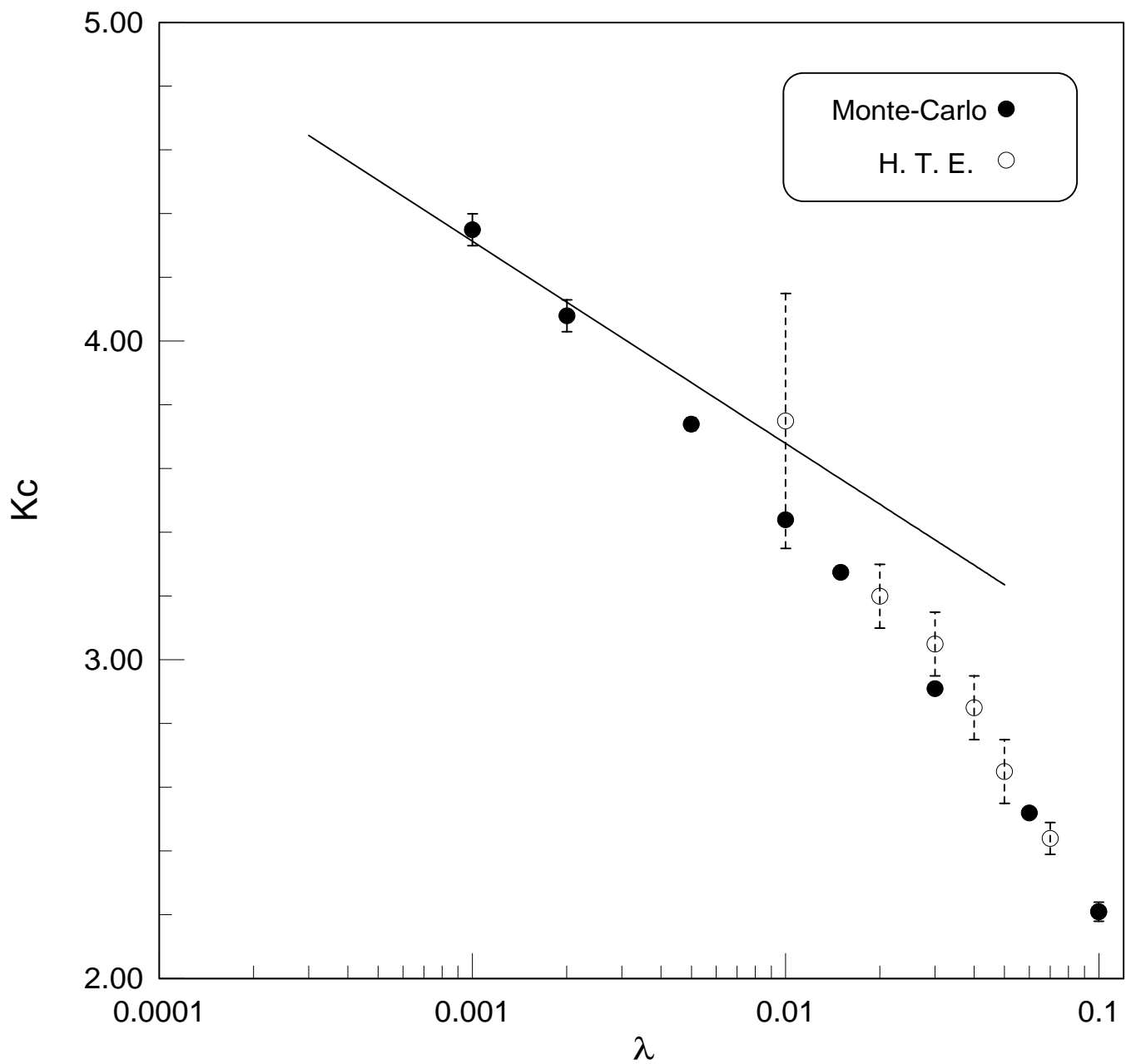
\item[{\bf Figure 5}]  Variation of the critical temperature with $\lambda $%
. The series results are represented by open circles and the Monte-Carlo
simulation by closed circles.

\begin{description}
\item[(a)]  The large anisotropy regime : we plot the inverse of the reduced
critical coupling $K_{c}^{-1}=\frac{2k_{B}T_{C}}{JS}$ as a function of the
anisotropy parameter. The slope of the linear behaviour, expected at large $%
\lambda $ corresponds to $K_{c\text{ Ising}}\simeq 1.46(1)$.

\item[(b)]  The weak anisotropy regime : the reduced critical coupling $%
K_{c} $ as a function of $\lambda $ in a logarithmic scale. The solid line
corresponds to the renormalisation group behaviour $K_{c}=-\frac{\sqrt{3}}{%
2\pi }\ln \lambda $ + const.$\;$The value of the constant is $2.41(2)$.
\end{description}

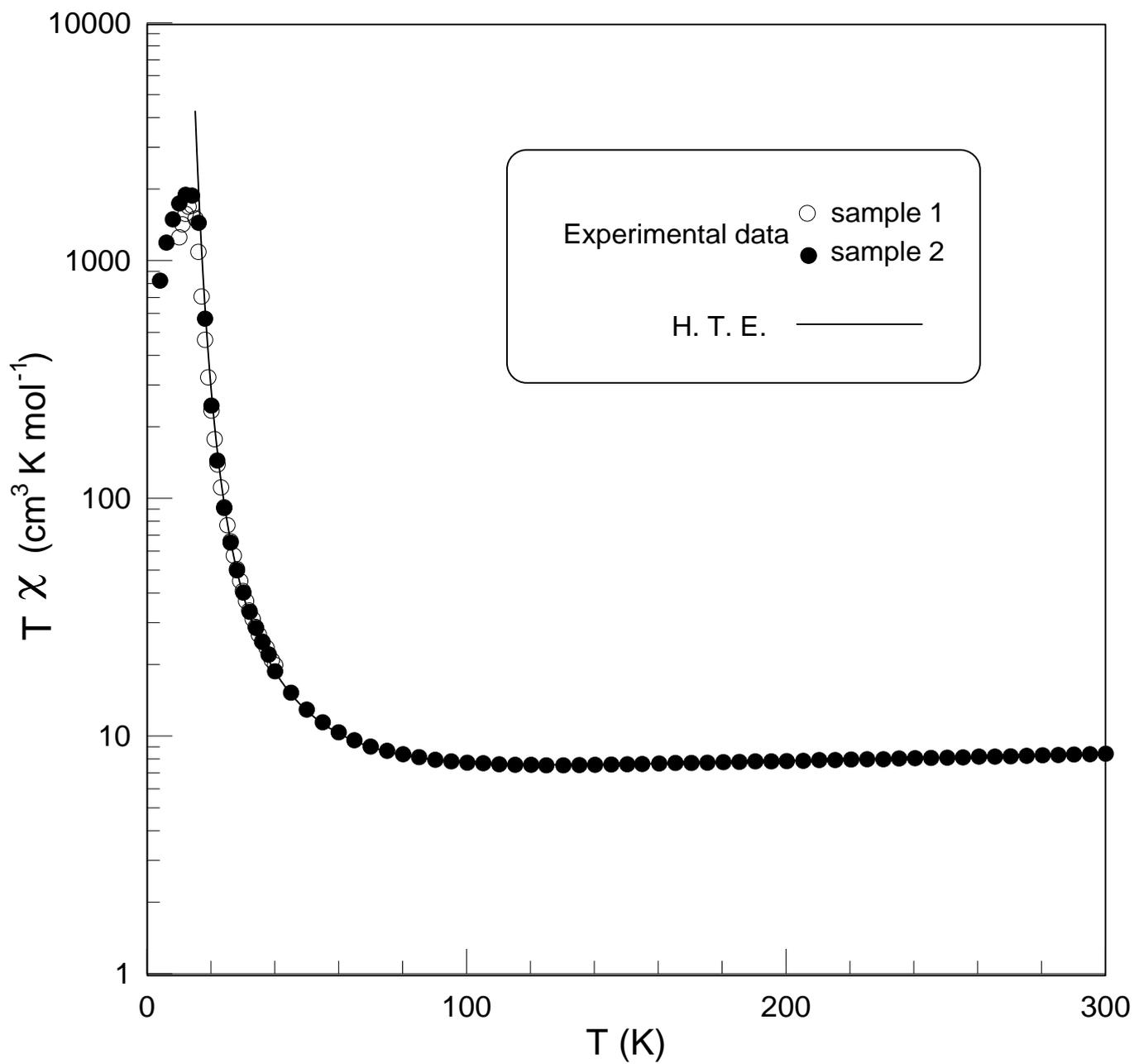
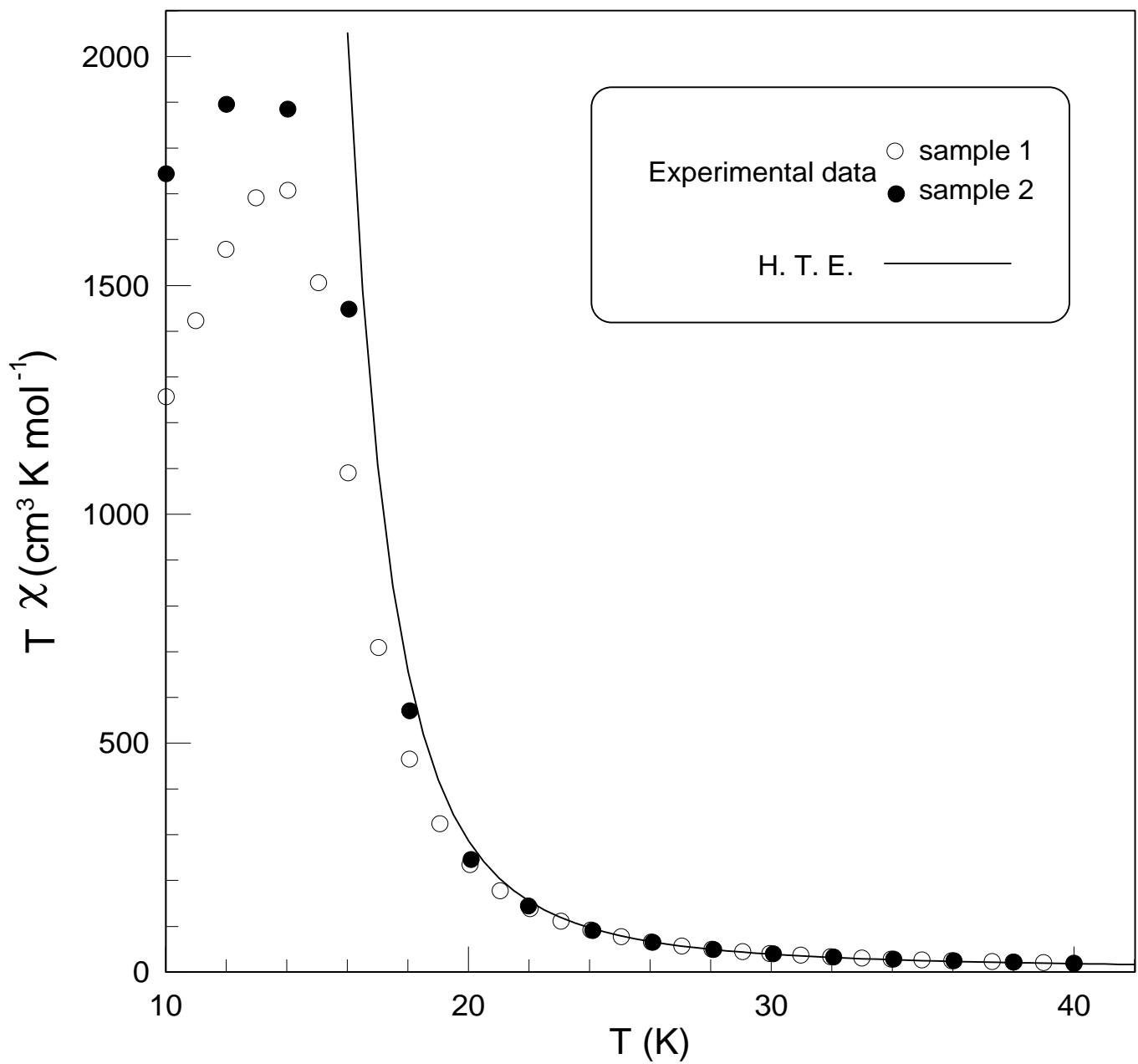
\item[{\bf Figure 6}]  Fit of the experimental data (closed circles for
sample 1, open circles for sample 2) for the total magnetic susceptibility.
The solid line corresponds to a Pad\'{e} approximant of the high temperature
series with the parameter : $J=45.5$ K, $g_{1}=2.0$, $g_{2}=2.14$, $\lambda
=.0005$;

\begin{description}
\item[(a)]  for the whole temperature range

\item[(b)]  for the critical region.\ The difference between the two samples
gives an estimation of the experimental errors in the critical region.
\end{description}
\end{description}

\end{document}